\documentclass[aip,jcp,reprint]{revtex4-1}
\usepackage{bm,graphicx,tabularx,array,booktabs,dcolumn,xcolor,microtype,multirow,amscd,amsmath,amssymb,amsfonts,physics,siunitx,tikz,wrapfig,enumitem}
\usepackage[version=4]{mhchem}
\usepackage[utf8]{inputenc}
\usepackage[T1]{fontenc}
\usepackage{txfonts}

\usepackage[normalem]{ulem}

\usepackage{natbib}
\AtBeginDocument{\nocite{achemso-control}}

\definecolor{hughgreen}{RGB}{0, 128, 0}

\newcommand{\edit}[1]{#1}

\usepackage{hyperref}
\hypersetup{
    colorlinks=true,
    linkcolor=blue,
    filecolor=blue,      
    urlcolor=blue,	
    citecolor=blue
}

\usepackage{caption} 
\usepackage{subcaption}
\usepackage{cleveref}

\usepackage{float} 

\newcommand{\sig}{\sigma}
\newcommand{\sigg}{\sig_g}
\newcommand{\sigu}{\sig_u}

\newcommand{\T}{\intercal}

\newcommand{\bc}{\bm{c}}

\newcommand{\bt}{\bm{t}}
\newcommand{\br}{\bm{r}}

\newcommand{\bH}{\bm{H}}
\newcommand{\bI}{\bm{I}}

\newcommand{\bQ}{\mathbf{Q}}

\newcommand{\bDelta}{\boldsymbol{\Delta}}

\newcommand{\etal}{\textit{et al.}} 

\newcommand{\cH}{\mathcal{H}}

\newcommand{\Nh}{N}

\newcommand{\UOX}{Physical and Theoretical Chemistry Laboratory, University of Oxford, South Parks Road, Oxford, OX1 3QZ, U.K.}

\begin{document}

\title{Energy Landscape of State-Specific Electronic Structure Theory}
\author{Hugh~G.~A.~Burton}
\email{hugh.burton@chem.ox.ac.uk}
\affiliation{\UOX}

\date{\today}

\begin{abstract}
\begin{wrapfigure}[11]{r}{0.4\textwidth}
    \flushleft
    \vspace{-0.4cm}
    \hspace{-1.65cm}
    \fbox{\includegraphics[width=0.4\textwidth]{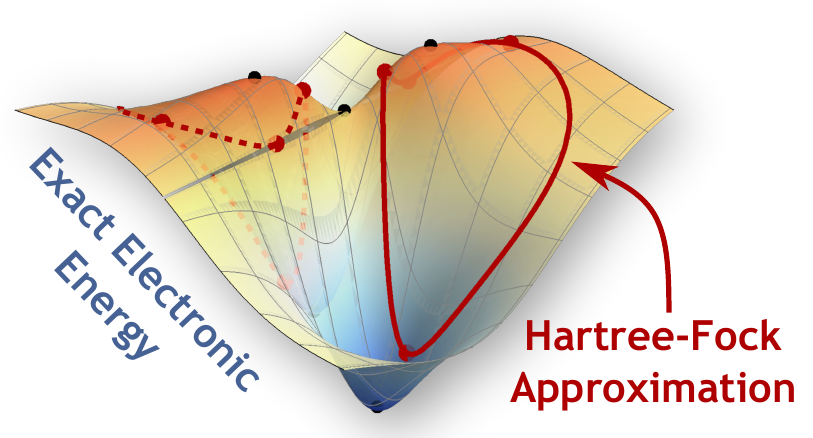}}
\end{wrapfigure}
State-specific approximations can provide an accurate representation of challenging electronic excitations by 
enabling relaxation of the electron density.
While state-specific wave functions are known to be local minima or saddle points of the approximate energy,
the \edit{global structure of the exact electronic energy} remains largely unexplored.
In this contribution, a \edit{geometric perspective on the exact electronic energy landscape} is introduced. 
On the exact energy landscape, ground and excited states form
stationary points constrained to the surface of a hypersphere and the corresponding Hessian index 
increases at each excitation level.
The connectivity between exact stationary points is investigated and the 
square-magnitude of the exact energy gradient is shown to be directly proportional to the
Hamiltonian variance.
The minimal basis Hartree--Fock and Excited-State Mean-Field representations of singlet \ce{H2} (STO-3G) are then used to
explore how the exact energy landscape controls the existence and properties of state-specific approximations.
In particular, approximate excited states correspond to constrained stationary points on the exact
energy landscape and their Hessian index also increases for higher energies. 
Finally, the properties of the  exact energy are used to derive the structure of the 
variance optimisation landscape and elucidate the challenges faced by variance optimisation
algorithms, including the presence of unphysical saddle points or maxima of the variance.
\end{abstract}

\maketitle

\raggedbottom

\section{Introduction}

Including wave function relaxation in state-specific approximations can provide an accurate 
representation of excited states where there is significant electron density rearrangement relative to the 
electronic ground state. 
This relaxation is particularly important for the description of charge
transfer,\cite{Barca2018,Liu2012,Jensen2018,Shea2018,Hait2021} 
core electron excitations,\cite{Hait2020a,Oosterbaan2018,Oosterbaan2020,Garner2020}
or Rydberg states with diffuse orbitals,\cite{CarterFenk2020,Shea2018,Clune2020}
\edit{and can be visualised using the eigenvectors of the difference density matrix for 
the excitation.\cite{Plasser2014a,Plasser2014b}}
In contrast, techniques based on linear response theory 
--- including time-dependent Hartree--Fock\cite{McLachlan1964} (TD-HF),
 time-dependent density functional theory\cite{Runge1984,Dreuw2005,Burke2005}  (TD-DFT),
configuration interaction singles\cite{Foresman1992,Dreuw2005}  (CIS),
and equation of motion coupled cluster theory\cite{Stanton1993,Krylov2008} (EOM-CC) ---
are evaluated using the ground-state orbitals, making a balanced treatment of the ground 
and excited states more difficult. \cite{Burke2005}
Furthermore, linear response methods are generally applied under the adiabatic approximation and are
limited to single excitations.\cite{Maitra2004,Burke2005} 
In principle, state-specific approaches can approximate both single and double 
excitations,\cite{Gilbert2008,Barca2018,Otis2020} although the open-shell character of single excitations requires a
multi-configurational approach.\cite{Shea2018,Hardikar2020,Zhao2020a,Zhao2020b,Ye2017}

Underpinning excited state-specific methods is the fundamental idea that ground-state wave functions
can also be used to describe an electronic excited state.
This philosophy relies on the existence of additional higher-energy mathematical solutions, which have been found
in Hartree--Fock (HF),\cite{Slater1951,Stanton1968,Fukutome1971,Fukutome1974a,Fukutome1974,Fukutome1975,Fukutome1973,Mestechkin1978,Mestechkin1979,Mestechkin1988,Davidson1983,Kowalski1998,Gilbert2008,Thom2008,Li2009a,Jimenez-Hoyos2011,Jimenez-Hoyos2014,Toth2016,Burton2018,Huynh2019,Lee2019,Burton2021} 
density functional theory (DFT),\cite{Theophilou1979,Perdew1985,Zarotiadis2020,Hait2020} 
multiconfigurational self-consistent field (MC-SCF),%
\cite{Olsen1982,Golab1983,Olsen1983,Golab1985,Angeli2003,Guihery1997,Evenhuis2011,Tran2019,Tran2020} 
and coupled cluster (CC) theory.\cite{PiecuchBook,Mayhall2010,Lee2019b,Marie2021a,Kossoski2021}
These multiple solutions correspond to higher-energy stationary points of a parametrised approximate energy function, 
including local energy minima, saddle points, or maxima.
\edit{It has long been known that the exact $k$-th excited state forms a saddle point of the energy 
with $k$ negative Hessian eigenvalues (where $k=0$ is the ground state).
These stationary properties have been identified using the exponential parametrisation of MC-SCF calculations%
\cite{Werner1981,Olsen1982,Golab1985,Olsen1983,Golab1983}
and can also be  derived using local expansions around an exact eigenstate.\cite{Bacalis2016,Bacalis2020}
However, questions remain about the global structure of the exact energy landscape and the 
connections between exact excited states. 
}

Multiple self-consistent field (SCF) solutions in HF or Kohn--Sham DFT are the most 
widely understood state-specific approximations. 
Their existence was first identified by Slater,\cite{Slater1951} and later characterised in detail by
Fukutome.\cite{Fukutome1971,Fukutome1974a,Fukutome1974,Fukutome1975,Fukutome1973}
Physically, these solutions appear to represent single-determinant approximations for 
excited states,\cite{Gilbert2008,Besley2009,Barca2014,Barca2018,Barca2018a}
or mean-field quasi-diabatic states.\cite{Thom2009,Jensen2018}
In the presence of strong electron correlation, multiple SCF solutions often break symmetries of the exact Hamiltonian
\cite{Davidson1983,Li2009a,Huynh2019,Jimenez-Hoyos2011,Jimenez-Hoyos2014,Lee2019}
and can disappear as the molecular geometry changes.\cite{Mestechkin1978,Mestechkin1979,Mestechkin1988,Burton2018}
The stability analysis pioneered by \v{C}i\v{z}ek and Paldus\cite{Cizek1967,Cizek1970,Thouless1960,Paldus1970} 
allows SCF solutions to be classified according to their Hessian index (the number of downhill orbital rotations). 
There are usually only a handful of low-energy SCF minima, connected by index-1 saddle points, while
symmetry-broken solutions form several degenerate minima that are connected by higher-symmetry saddle points.\cite{Burton2021}
At higher energies, stationary points representing excited states generally become higher-index saddle points of 
the energy.\cite{Perdew1985,Dardenne2000,Burton2021}

Recent interest in locating higher-energy SCF solutions has led to several new approaches including: 
modifying the iterative SCF approach with orbital occupation constraints\cite{Gilbert2008,Barca2018} or level-shifting;\cite{CarterFenk2020}
second-order direct optimisation of higher-energy stationary points;\cite{Levi2020,Levi2020a}
minimising an alternative functional such as the variance\cite{Ye2017,Ye2019,Shea2017,Shea2018,Cuzzocrea2020}  
or the square-magnitude of the energy gradient.\cite{Hait2020}
\edit{The success of these algorithms depends on the structure of the approximate energy landscape, the 
stationary properties of excited states, and the quality of the initial guess.
In principle, the approximate energy landscape is determined by the relationship between an approximate wave function and the 
exact energy landscape.
However,  the nature of this connection has not been widely investigated.}

Beyond single-determinant methods, state-specific approximations using multi-configurational wave functions 
have been developed to describe open-shell or statically correlated excited states, including
MC-SCF,\cite{Olsen1982,Golab1985,Olsen1983,Golab1983,Tran2019,Tran2020} 
excited-state mean-field (ESMF) theory,\cite{Shea2018,Hardikar2020,Zhao2020a,Zhao2020b}
half-projected HF,\cite{Ye2019} or multi-Slater-Jastrow functions.\cite{Dash2019,Cuzzocrea2020,Dash2021} 
\edit{The additional complexity of these wave functions compared to a single determinant has led to the use
of direct second-order optimisation algorithms\cite{Olsen1982,Golab1985,Olsen1983,Golab1983,Werner1981}
or, more recently,
methods based on variance optimisation.\cite{Messmer1969,Shea2017,PinedaFlores2019}}
Variance optimisation exploits the fact that both ground and excited states form minima of the 
Hamiltonian variance $\mel*{\Psi}{(\cH - E)^2}{\Psi}$,\cite{MacDonald1934}
and thus excited-states can be identified using downhill minimisation techniques.
\edit{Alternatively, the folded-spectrum method uses an objective function with the form $\mel*{\Psi}{(\cH - \omega)^2}{\Psi}$ to 
target the state with energy closest to $\omega$.\cite{Wang1994}}
These approaches are particularly easy to combine with stochastic methods such as 
variational Monte--Carlo\cite{Umrigar2005,PinedaFlores2019,Cuzzocrea2020} (VMC) and have 
been proposed as an excited-state extension of variational quantum eigensolvers.\cite{Zhang2020,Zhang2021}
However, variance optimisation is prone to convergence issues that include
drifting away from the intended target state,\cite{Cuzzocrea2020,Otis2020} and very little is known about the 
properties of the variance optimisation landscape, or its stationary points.

In my opinion, our limited understanding about the relationship between exact and approximate 
state-specific solutions arises because exact electronic structure is traditionally 
viewed as a matrix eigenvalue problem, while state-specific approximations are considered as higher-energy stationary 
points of an energy landscape.
The energy landscape concept is more familiar to theoretical chemists in the context
of a molecular potential energy surface,\cite{Born1927} where 
local minima correspond to stable atomic arrangements and index-1 saddles can be interpreted as 
reactive transition states.\cite{Murrell1968,WalesBook}  
To bridge these concepts, this article introduces a fully geometric perspective on 
exact electronic structure within a finite Hilbert space.
In this representation, ground and excited states form stationary points of an energy landscape constrained 
to the surface of a unit hypersphere.
Analysing the differential geometry of this landscape reveals the stationary properties of ground and excited states
and the pathways that connect them. 
Furthermore, the square-magnitude of the exact gradient is shown to be directly proportional to the Hamiltonian
variance, allowing the structure of the exact variance optimisation landscape to be derived.
Finally, the relationship between approximate wave functions and the exact energy or variance is 
explored, revealing how the stationary properties of state-specific solutions are controlled
by the structure of the exact energy landscape. 

Throughout this work, key concepts are illustrated using the electronic singlet states
of \ce{H2}.
While this minimal model is used to allow visualisation of the exact energy landscape,
the key conclusions are mathematically general and can be applied to any number of electrons or basis functions.
Unless otherwise stated, all results are obtained with the STO\nobreakdash-3G basis set\cite{Hehre1969} 
using Mathematica 12.0\cite{Mathematica} and are available in an accompanying notebook available for 
download from \href{https://doi.org/10.5281/zenodo.5615978}{https://doi.org/10.5281/zenodo.5615978}.
Atomic units are used throughout.

\section{Exact Electronic Energy Landscape}
\label{sec:ExactLandscape}

\subsection{Traditional eigenvalue representation}

For a finite $\Nh$-dimensional Hilbert space, the exact electronic wave function can 
be represented using a full configuration interaction (FCI) expansion 
constructed from a linear expansion of orthogonal Slater determinants as
\begin{equation}
\ket{\Psi} = \sum_{I=1}^{\Nh} c_{I} \ket{\Phi_I}.
\end{equation}
This expansion is invariant to the particular choice of orthogonal Slater determinants $\ket{\Phi_I}$, but the 
set of all excited configurations from a self-consistent HF determinant is most commonly used.\cite{SzaboBook}
Normalisation of the wave function introduces a constraint on the expansion coefficients 
\begin{equation}
\sum_{I=1}^{\Nh} \abs{c_{I}}^2 = 1,
\label{eq:ExactNorm}
\end{equation}
while the electronic energy is given by the Hamiltonian expectation value 
\begin{equation}
E = \mel{\Psi}{\cH}{\Psi} = \sum_{I,J=1}^{\Nh} c_{I}^{*} \mel{\Phi_I}{\cH}{\Phi_J} c_{J}^{\vphantom{*}}.
\end{equation}
The optimal coefficients are conventionally identified by solving the secular equation
\begin{equation}
\sum_{J=1}^{\Nh} \mel{\Phi_I}{\cH}{\Phi_J} c_{J}^{\vphantom{*}} = E\, c_{I},
\end{equation}
giving the exact ground- and excited-state energies as the eigenvalues of the Hamiltonian matrix $H_{IJ} = \mel{\Phi_I}{\cH}{\Phi_J}$.

\subsection{Differential geometry of electronic structure theory}

\begin{figure}[t]
\includegraphics[width=\linewidth]{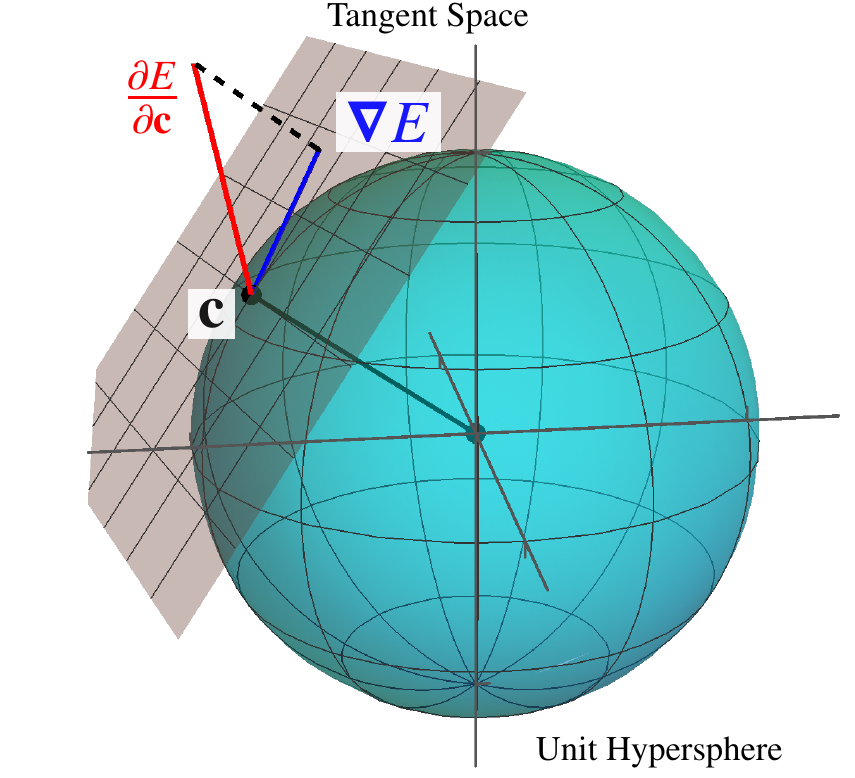}
\caption{Geometric relationship between the position vector $\bc$ (black) for a wave function constrained to the 
unit hypersphere, the global energy gradient in the full Hilbert space [red; Eq.~\eqref{eq:GlobalGradient}], 
and the local gradient in the tangent space [blue; Eq.~\eqref{eq:LocalGradient}]}  \label{fig:localGradient}
\end{figure}

In what follows, the wave function and Hamiltonian are assumed to be real-valued, although the key conclusions
can be extended to complex wave functions.
A geometric energy landscape for exact electronic structure can be constructed by representing 
the wave function $\ket{\Psi}$ as an $\Nh$-dimensional  vector $\bc \in \mathbb{R}^\Nh$ with coefficients
\begin{equation}
\bc = 
\begin{pmatrix}
	c_{1}, \cdots, c_{\Nh}
\end{pmatrix}^{\intercal}.
\label{eq:VectorForm}
\end{equation}
The energy is then defined by the quadratic expression
\begin{equation}
E(\bc) = \bc^\T \bH \bc^{\vphantom{\T}},
\end{equation}
where $\bH$ represents the Hamiltonian matrix in the orthogonal basis of 
Slater determinants with elements $H_{IJ}~=~\mel{\Phi_I}{\cH}{\Phi_J}$.\cite{SzaboBook}
Normalisation of the wave function is geometrically represented as 
\begin{equation}
	\bc^{\T} \bc^{\vphantom{\T}} = 1
\end{equation}
and requires the coefficient vector $\bc$ to be constrained to a unit hypersphere of dimension $(\Nh-1)$ 
embedded in the full $\Nh$-dimensional space (see Fig.~\ref{fig:localGradient}).
In this representation, optimal ground and excited states are stationary points of the 
electronic energy constrained to the surface of this unit hypersphere. 

The stationary conditions are obtained by applying the framework
of differential geometry under orthogonality constraints, as described in  Ref.~\onlinecite{Edelman1998}.
In particular, a stationary point requires that the global gradient of the energy in the full Hilbert space, 
given by the vector
\begin{equation}
\frac{\partial E}{\partial \bc} = 2\, \bH \bc,
\label{eq:GlobalGradient}
\end{equation}
has no component in the tangent space to the hypersphere.
%
At a point $\bc$ on the surface of the hypersphere, tangent vectors $\bDelta$  satisfy 
the condition\cite{Edelman1998}
\begin{equation}
	\bDelta^{\T} \bc + \bc^{\T} \bDelta = 0.
\end{equation}
The orthogonal basis vectors that span this $(\Nh-1)$-dimensional tangent space form the columns of 
a projector into the local tangent basis,\cite{Savas2010}  
denoted $\bc_{\bot} \in \mathbb{R}^{\Nh \times (\Nh-1)}$ .
\edit{Note that $\bc_{\bot}$ forms a \textit{matrix} whose columns span the tangent space, while $\bc$ is
a \textit{vector} representing the current position.}
The corresponding projectors satisfy the completeness condition
\begin{equation}
\bc \bc^{\T} + \bc_{\bot}^{\vphantom{\T}} \bc_{\bot}^{\T} = \bI_{\Nh},
\label{eq:ProjectorExpansion}
\end{equation}
\edit{where $\bI_{\Nh}$ is the $\Nh$-dimensional identity matrix},
and span disjoint vector spaces such that
\begin{equation}
\bc \bc_{\bot}^{\T}  = \bc_{\bot}  \bc ^{\T} = \bm{0}.
\label{eq:ProjectorExpansion2}
\end{equation}
The constrained energy gradient is then obtained by 
projecting the global gradient Eq.~\eqref{eq:GlobalGradient} into the  tangent space to give
\begin{equation}
	\grad E = \bc_{\bot}^\T \frac{\partial E}{\partial \bc} = 2\, \bc_{\bot}^\T \bH \bc,
\label{eq:LocalGradient}
\end{equation}
with constrained stationary points satisfying  $\grad E = \bm{0}$.
Figure~\ref{fig:localGradient} illustrates this geometric relationship between the unit hypersphere, 
the exact tangent space, the global gradient in the full Hilbert space, and the local gradient in the tangent space.

The stationary condition $\grad E = \bm{0}$ requires that the global gradient Eq.~\eqref{eq:GlobalGradient} has no 
component in the tangent space. 
Therefore, it can only be satisfied if $\bH \bc$ is (anti)parallel to the position vector $\bc$.
This condition immediately recovers the expected eigenvector expression $\bH \bc_k = E_k \bc_k$, where the 
eigenvalue $E_k$ is the exact energy of the $k$-th excited state with coefficient vector $\bc_k$.
As a result, each exact eigenstate is represented by two stationary points on the hypersphere that are related
by a sign-change in the wave function (i.e., $\pm \bc_k$ or, equivalently, $\pm \ket{\Psi_k}$).
There are no other stationary points on the exact landscape.

The exact hypersphere can be compared to the constraint surface in HF theory, 
where the occupied orbitals represent the current position on a Grassmann manifold and occupied-virtual orbital rotations 
define the tangent basis vectors.\cite{Voorhis2002,Edelman1998}
In HF theory, the global gradient is given by $2 \bm{F}\bm{C}_\text{occ}$, 
where $\bm{F}$ is the Fock matrix and 
$\bm{C}_\text{occ(vir)}$ are the occupied (virtual) orbital coefficients. 
Projection into the Grassmann tangent space then yields the local gradient as 
$2 \bm{C}_\text{vir}^{\T} \bm{F}\bm{C}_\text{occ}^{\vphantom{\T}}$, which 
corresponds to the virtual--occupied block of the Fock matrix in the molecular orbital (MO)
basis.\cite{Douady1980,Voorhis2002,Chaban1997,Burton2021}
Therefore, in both HF theory and the exact formalism, optimisation of the energy requires the off-diagonal 
blocks of an effective Hamiltonian matrix to become zero such that the ``occupied--virtual'' coupling between orbitals
or many-particle states vanishes.

\subsection{Properties of exact stationary points}
\label{subsec:exactProperties}

\begin{figure*}
\includegraphics[width=\linewidth]{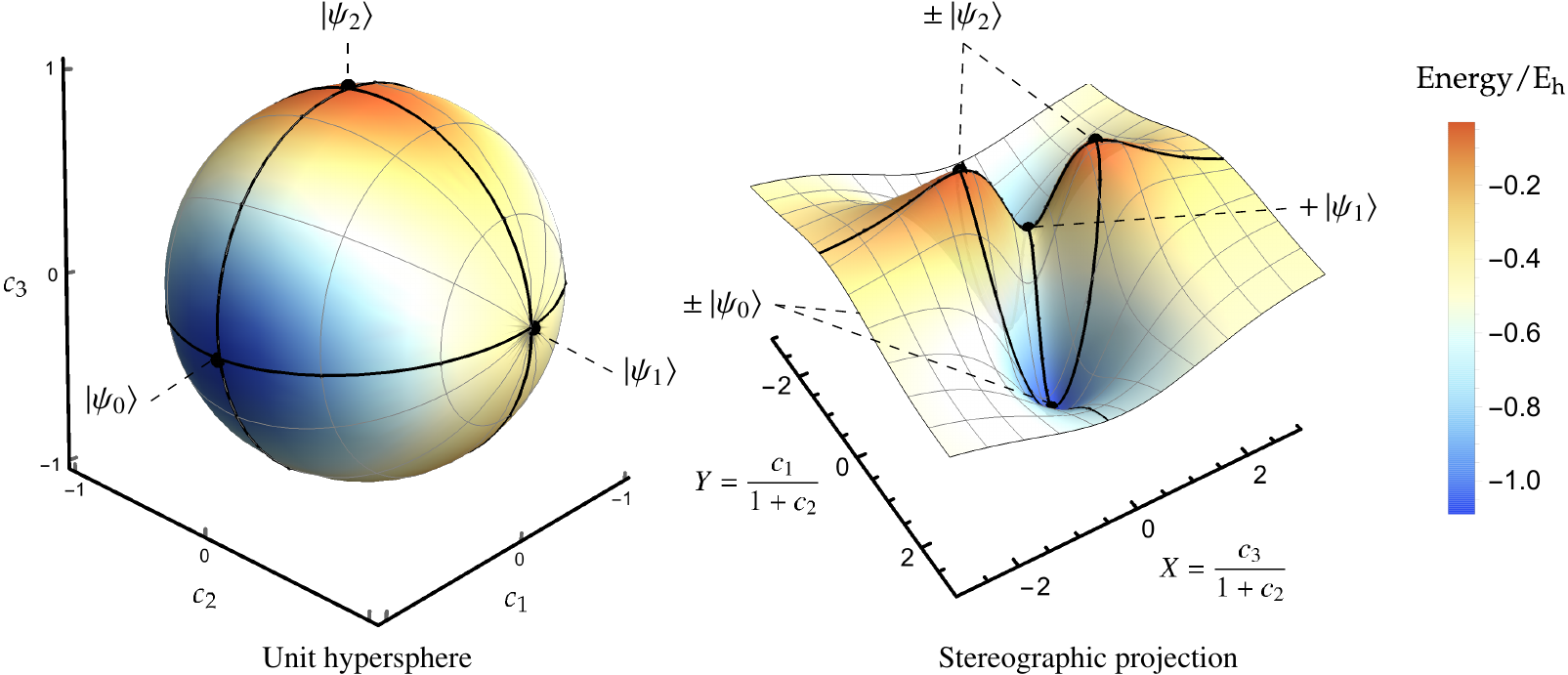}
\caption{\textit{Left:} Exact singlet electronic energy for \ce{H2} (STO-3G) at a bond length of $\SI{2}{\bohr}$ constrained 
to the unit hypersphere.
\textit{Right:} Stereographic projection of the exact singlet energy using Eq.~\eqref{eq:stereo}. 
Ground and excited states correspond to stationary points of the energy (black dots). 
The $-\ket{\Psi_1}$ state is at infinity in this representation.
Each pair of stationary points is directly connected by a gradient extremal (black line) where the gradient is an eigenvector of the Hessian.
}
\label{fig:fciTopology}
\end{figure*}

Stationary points on an energy landscape can be characterised as either minima, 
index-$k$ saddles, or maxima, 
depending on the number of downhill directions (or negative eigenvalues of the Hessian matrix).
Following Ref.~\onlinecite{Edelman1998}, the analytic Hessian $\bQ$ of the exact energy constrained to
the hypersphere is given by
\begin{equation}
\bQ = \frac{\partial^2 E}{\partial \bc^2} - \qty(\frac{\partial E}{\partial \bc})^\T \bc.
\end{equation}
Taking the global second derivative in the full Hilbert space 
\begin{equation}
\frac{\partial^2 E}{\partial \bc^2}  = 2 \bH
\end{equation}
and exploiting the relationship
\begin{equation}
 \qty(\frac{\partial E}{\partial \bc})^\T \bc = 2\, \bc^\T \bH \bc = 2 E
\end{equation}
allows the Hessian to be expressed as a shifted and rescaled Hamiltonian matrix
\begin{equation}
\bQ = 2\qty(\bH - E\,\bI_\Nh).
\label{eq:GlobalHessian}
\end{equation}
Projecting into the space spanned by the tangent vectors then gives the constrained 
local Hessian as an $(\Nh-1)\times(\Nh-1)$ matrix defined as
\begin{equation}
\widetilde{\bQ} = 2\,\bc_{\bot}^{\T} \qty(\bH - E\,\bI_{\Nh})\,\bc_{\bot}^{\vphantom{\T}} ,
\label{eq:LocalHessian}
\end{equation}
\edit{where it should be remembered that $\bc_\bot$ is an $N\times(N-1)$ matrix.}
This expression for the Hessian of a CI wave function is also obtained with the exponential 
transformation used in second-order MC-SCF approaches.\cite{Olsen1983,RoosBook}

The constrained Hessian allows the properties of exact stationary points to be recovered.
Consider a constrained stationary point $\bc_k$ corresponding to an exact 
eigenstate $\ket{\Psi_k}$ with energy $E_k$.
To satisfy the completeness condition Eq.~\eqref{eq:ProjectorExpansion}, the tangent basis vectors
at this point must correspond to the remaining $(\Nh-1)$ exact eigenstates.
The local Hessian $\widetilde{\bQ}$ is then simply the full Hamiltonian shifted by $E_k$ and projected into 
the basis of these $(\Nh-1)$ exact eigenstates.
As a result, the Hessian eigenvalues $\lambda_i$ are directly proportional to the excitation energies, giving
\begin{equation}
\lambda_i =2(E_{i} - E_{k}) = 2\Delta E_{ik},
\end{equation}
where $E_i$ are the exact energies with $i\neq k$.
Furthermore, the corresponding eigenvectors coincide with the position vectors $\bc_i$ representing the 
remaining exact eigenstates $\ket{\Psi_i}$.
Remarkably, this means that only the energy and second derivatives 
at an exact stationary point are required to deduce the electronic energies of the  \textit{entire} system.
In other words, the full electronic energy spectrum is encoded in the local structure of the 
energy landscape around a single stationary point.

Now, at the stationary point $\bc_k$ with energy $E_k$, the number of exact eigenstates 
that are lower in energy is equal to the excitation level $k$.
The number of negative eigenvalues $\lambda_i = 2 \Delta E_{ik}$ is then equivalent to the excitation level. 
Significantly, there are only two minima on the exact energy landscape 
corresponding to positive and negative sign-permutations of the exact ground state, i.e.\ $\pm \ket{\Psi_0}$,
\edit{and no higher-energy local minima.}
The $k$-th excited state forms a pair of index-$k$ saddle points that are also related by 
a sign-change in the wave function. 
\edit{The saddle-point nature of exact excited states was previously derived in the context of MC-SCF 
theory,\cite{Olsen1982,Olsen1983,Golab1985,Golab1983} and has been described by Bacalis using local expansions
around an excited state.\cite{Bacalis2020}
In fact, the Hessian index has been suggested as a means of targeting and characterising a 
particular MC-SCF excited state.\cite{Golab1983,Olsen1983}
In contrast, here the stationary properties of exact excited states have been derived using only the
differential geometry of functions under orthogonality constraints.
As will be shown later, this differential geometry also reveals the global structure of the energy 
landscape and the connections between exact eigenstates.
}

These properties also apply within a particular symmetry subspace. 
For example, the first excited state of a given symmetry is an index-1 saddle  on the energy landscape 
projected into the corresponding symmetry subspace, but may be a higher-index saddle on the full energy landscape. 
\edit{For a pair of degenerate eigenstates, the corresponding stationary points will have a zero Hessian 
eigenvalue and the corresponding eigenvector will interconvert the two states. 
Therefore, degenerate eigenstates form a flat continuum of stationary points on the exact energy landscape 
and any linear combination of the two states must also be a stationary point of the energy.}

In Fig.~\ref{fig:fciTopology}, the structure of the exact energy landscape is illustrated for the 
 singlet states of \ce{H2} at a bond length of $\SI{2}{\bohr}$ using the STO-3G basis set.\cite{Hehre1969}
An arbitrary spin-pure singlet wave function can be constructed as a linear combination 
of singlet configuration state functions to give
\begin{equation}
\ket{\Psi} = c_1 \abs*{\sigg \bar{\sigg}} + \frac{c_2}{\sqrt 2} \qty(\abs*{\sigg \bar{\sigu}} + \abs*{\sigu \bar{\sigg}}) + c_3 \abs{\sigu \bar{\sigu}}.
\label{eq:genH2wfn}
\end{equation}
Here, $\sigg$ and $\sigu$ are the symmetry-adapted MOs,
and the absence (presence) of an overbar indicates an occupied high-spin (low-spin) orbital. 
A stereographic projection  centred on $(c_1,c_2,c_3) = (0,1,0)$ is used to 
highlight the topology of the energy landscape (Fig.~\ref{fig:fciTopology}: right panel), 
with new coordinates $X$ and $Y$ defined as
\begin{align}
X = \frac{c_3}{1+c_2}
\quad\text{and}\quad
Y = \frac{c_1}{1+c_2}.
\label{eq:stereo}
\end{align}
The ground and excited states (black dots) form stationary points constrained to the hypersphere 
(Fig.~\ref{fig:fciTopology}: left panel) with the global minima
representing the ground state, index-1 saddles representing the first excited state, and the global maxima 
representing the second excited state (Fig.~\ref{fig:fciTopology}: right panel).
At the index-1 saddle, the downhill directions connect the two sign-permutations of 
the ground-state wave function, while 
the two uphill directions connect sign-permutations of the second excited singlet state.


\subsection{Gradient extremals on the electronic energy surface}
\label{subsec:GradExtremal}

Energy landscapes can also be characterised by the pathways that connect stationary points. 
For molecular potential energy surfaces, pathways can be interpreted as
reaction trajectories between stable molecular structures, with saddle points representing 
reactive transition states.\cite{WalesBook}
However, unlike stationary points, these pathways do not have a unique mathematical definition. 
On the exact electronic  energy landscape, gradient extremals represent the most 
obvious pathways between stationary points.
A gradient extremal is defined as a set of points where the gradient is either maximal or minimal 
along successive energy-constant contour lines.\cite{Hoffman1986}
For the contour line with energy $E_\text{c}$, these points can be identified by the 
constrained optimisation\cite{SzaboBook}
\begin{equation}
\begin{split}
\frac{\partial }{\partial \bc}\qty[ \abs{\grad E}^2 - 2 \lambda \qty(E - E_c ) ] 
=2 \qty[ \widetilde{\bQ}\cdot \grad E - \lambda \grad E ] = 0.
\end{split}
\label{eq:SqGradCondition1}
\end{equation}
Therefore, gradient extremals are pathways where the local gradient is an eigenvector of the 
local Hessian, i.e.
\begin{equation}
\widetilde{\bQ}(\bc)\, \grad E(\bc) = \lambda(\bc) \grad E(\bc).
\label{eq:SqGradCondition2}
\end{equation}
These pathways propagate away from each stationary point along the ``normal mode'' eigenvectors of the Hessian
and provide the softest or steepest ascents from a minimum.\cite{Hoffman1986}

On the exact FCI landscape, gradient extremals directly connect each pair of stationary points, 
as illustrated by the black lines in Fig.~\ref{fig:fciTopology}.
Each extremal corresponds to the geodesic connecting the two stationary points along the 
surface of the hypersphere. 
The wave function along these pathways is only a linear combination of the  
two eigenstates at each end of the path.
Therefore, gradient extremals provide a well-defined route along which a ground-state wave function can 
be continuously evolved into an excited-state wave function, or vice-versa.
Furthermore, as a gradient extremal moves from the lower-energy to the higher-energy stationary point, 
the corresponding Hessian eigenvalue $\lambda(\bc)$ changes from positive to negative.
This leads to an inflection point with $\lambda(\bc) = 0$ exactly halfway along each gradient extremal
where the wave function is an equal combination of the two exact eigenstates.
These inflection points are essential for understanding the structure of the variance optimisation 
landscape in Section~\ref{subsec:SquaredGradientVariance}.

\subsection{Structure of the exact variance landscape}
\label{subsec:SquaredGradientVariance}

\begin{figure*}
\includegraphics[width=\linewidth]{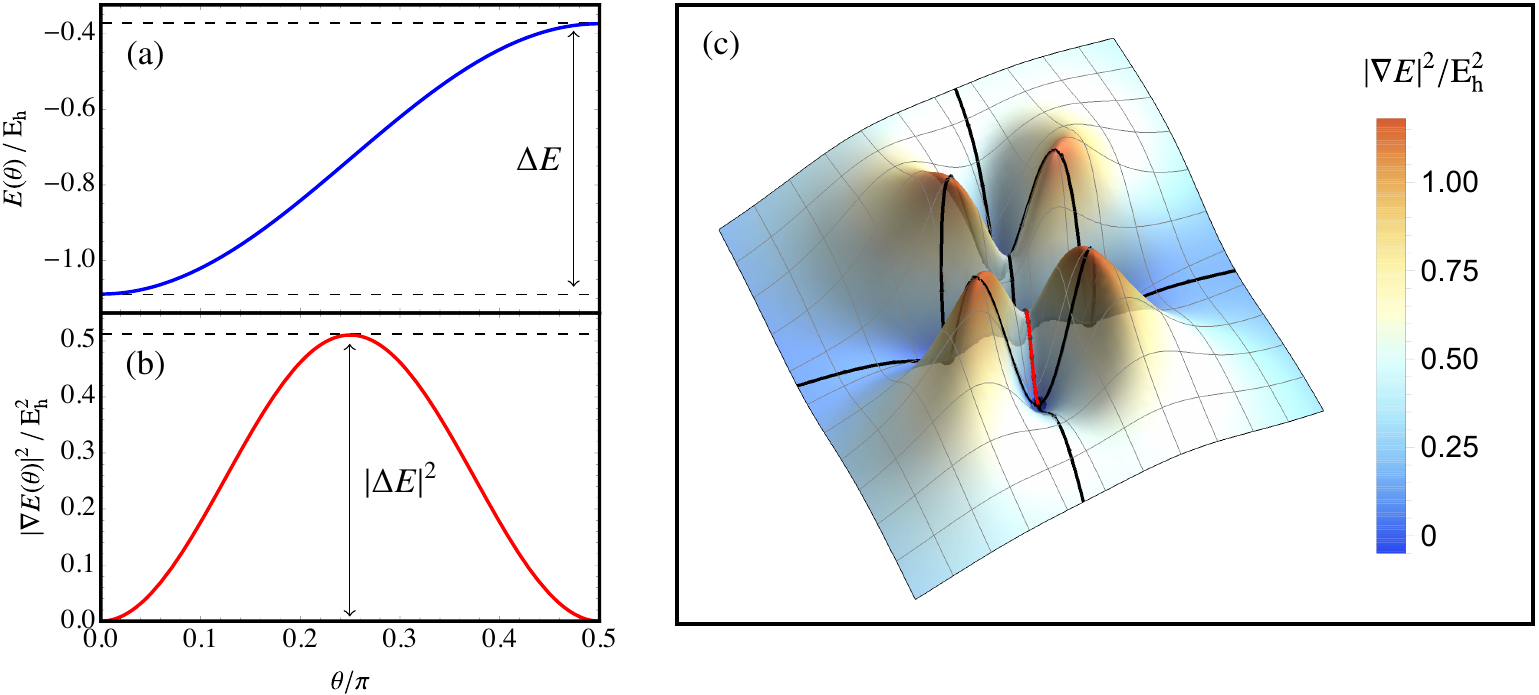}
\caption{Energy square-gradient for the singlet states of \ce{H2} (STO-3G) at a bond length of $\SI{2}{\bohr}$. 
(a) Energy along the gradient extremal connecting the ground and first excited singlet state.
(b) Square-gradient of the energy along the gradient extremal connecting the ground and first excited singlet state.
(c) Square-gradient landscape for singlet wave functions in \ce{H2} (STO-3G), with gradient extremals (black) connecting the physical minima. The gradient extremal plotted in (b) is highlighted in red.
}
\label{fig:fciVariance}
\end{figure*}

The accuracy of a general point on the exact energy landscape can
be assessed using the square-magnitude of the local gradient,  defined using Eq.~\eqref{eq:LocalGradient} as
\begin{equation}
\abs{\grad E(\bc)}^2 =  4\, \bc^\T \bH \bc_{\bot}^{\vphantom{\T}} \bc_{\bot}^\T \bH \bc.
\label{eq:GradMag1}
\end{equation}
Since all stationary points are minima of the squared-gradient
with $\abs{\grad E(\bc)}^2=0$, minimising this objective function has been proposed 
as a way of locating higher-index saddle points in various contexts.\cite{Broderix2000,Angelani2000,Hait2020}
For the electronic structure problem, exploiting the relationship between the tangent- 
and normal-space projectors [Eq.~\eqref{eq:ProjectorExpansion}] allows Eq.~\eqref{eq:GradMag1} to be 
expressed using only the position vector $\bc$ as
\begin{equation}
\abs{\grad E}^2 =  4\,  \bc^\T \bH \qty(\bI_{\Nh} - \bc \bc^\T) \bH \bc.
\label{eq:GradMag2}
\end{equation}
Further expanding this expression gives
\begin{equation}
\abs{\grad E}^2 =  4\qty( \bc^\T \bH^2  \bc -  \qty(\bc^\T \bH \bc)^2),
\label{eq:GradMag3}
\end{equation}
which can be recognised as four times the Hamiltonian variance, i.e.\
\begin{equation}
\abs{\grad E}^2  = 4\mel*{\Psi}{(\cH - E)^2}{\Psi}.
\end{equation}
While variance optimisation has previously inspired the development of excited-state variational
principles,\cite{Messmer1969,Shea2017,PinedaFlores2019,Cuzzocrea2020,Ye2017,Ye2019} these approaches
have generally been motivated by the fact that exact eigenstates of $\cH$ also have zero variance.
In contrast, the relationship between the variance and $\abs{\grad E}^2$ provides a purely geometric 
motivation behind searching for excited states in this way, derived from the structure of the exact energy landscape.
Hait and Head-Gordon alluded to a relationship of this type by noticing similarities 
between the equations for SCF square-gradient minimisation and optimising the SCF variance.\cite{Hait2020} 

By connecting the squared-gradient of the energy to the variance, the exact energy 
landscape can be used to deduce the structure of the variance optimisation landscape.
Exact eigenstates form minima on the square-gradient landscape with $\abs{\grad E}^2 = 0$.
However, the square-gradient  can have additional non-zero stationary points 
corresponding to local minima, higher-index saddle points, or local maxima.\cite{Doye2002,Doye2003}
These ``non-stationary'' points do not represent stationary points of the energy, but 
they  provide important information about the structure of the square-gradient landscape away from exact 
eigenstates. 
In particular,
a stationary point of 
$\abs{\grad E}^2$ with $\grad E \neq \bm{0}$ can only occur when the local gradient is an eigenvector
of the Hessian with a zero eigenvalue, $\widetilde{\bQ}(\bc)\, \grad E(\bc) = \bm{0}.$\cite{Doye2002,Doye2003} 
Non-stationary points therefore occur on the gradient extremals
described in Section~\ref{subsec:GradExtremal} and correspond to the inflection points 
exactly halfway between each pair of eigenstates.

Since gradient extremals only connect two eigenstates, the value of $\abs{\grad E}^2$ at these
non-stationary points can be obtained by parametrising the wave function as 
 \begin{equation}
\ket{\Psi(\theta)} = \cos \theta \ket*{\Psi_i} + \sin \theta \ket*{\Psi_j}.
\end{equation}
The energy and square-gradient are then given by 
 \begin{subequations}
\begin{align}
E(\theta) &= E_i + \Delta E_{ji}\, \sin^2 \theta,
\\
\abs{\grad E(\theta)}^2 &= \Delta E_{ji}^2\, \sin^2 2\theta,
\end{align}
\end{subequations}
where $0 \le \theta \le \pi / 2$ and  $\Delta E_{ji} = E_j - E_i$.
There are only two stationary points of the energy along each pathway, corresponding to 
$\abs{\grad E}^2 = 0$ at $\theta = 0$ and $\pi / 2$, as illustrated for the gradient extremal 
connecting the ground and first excited singlet states of \ce{H2} (STO-3G) in Fig.~\ref{fig:fciVariance}a. 
In contrast, the square-gradient has an additional stationary point at the inflection point $\theta = \pi/4$ with 
$\abs{\grad E}^2 = \Delta E_{ji}^2$ (see Fig.~\ref{fig:fciVariance}b). 
This point corresponds to an unphysical maximum of 
$\abs{\grad E}^2 $ along the gradient extremal and the height of this barrier depends on the square of the energy 
difference between the two states. 
Therefore,  the exact square-gradient (or variance) landscape
contains exact minima separated by higher-variance stationary points that form barriers at the inflection points of the energy, 
with the height of each barrier directly proportional to 
the square of the energy difference between the connected eigenstates.
The lowest square-gradient barrier always connects states that are adjacent in energy
to form an index-1 saddle point, while barriers connecting states that are not adjacent in energy 
form higher-index saddle points of $\abs{\grad E}^2$. %
This structure of the exact square-gradient landscape is illustrated for the singlet states of 
\ce{H2} in Fig.~\ref{fig:fciVariance}c.

Connecting the variance to the exact energy square-gradient reveals that the general structure of the 
variance optimisation landscape is universal and completely determined by the energy difference between 
exact eigenstates.
Systems with very small energy gaps will have low barriers between exact variance minima,
while systems with well-separated energies will have high-variance barriers.
\edit{This structure may also play a role in explaining the convergence behaviour of variance minimisation 
approaches, as discussed in Section~\ref{sec:Discussion}.}

\section{Understanding Approximate Wave Functions}
\label{sec:ApproximateMethods}

\subsection{Differential geometry on the exact energy landscape}
\label{subsec:RelationshipToExactEnergy}

\begin{figure*}
\includegraphics[width=\linewidth]{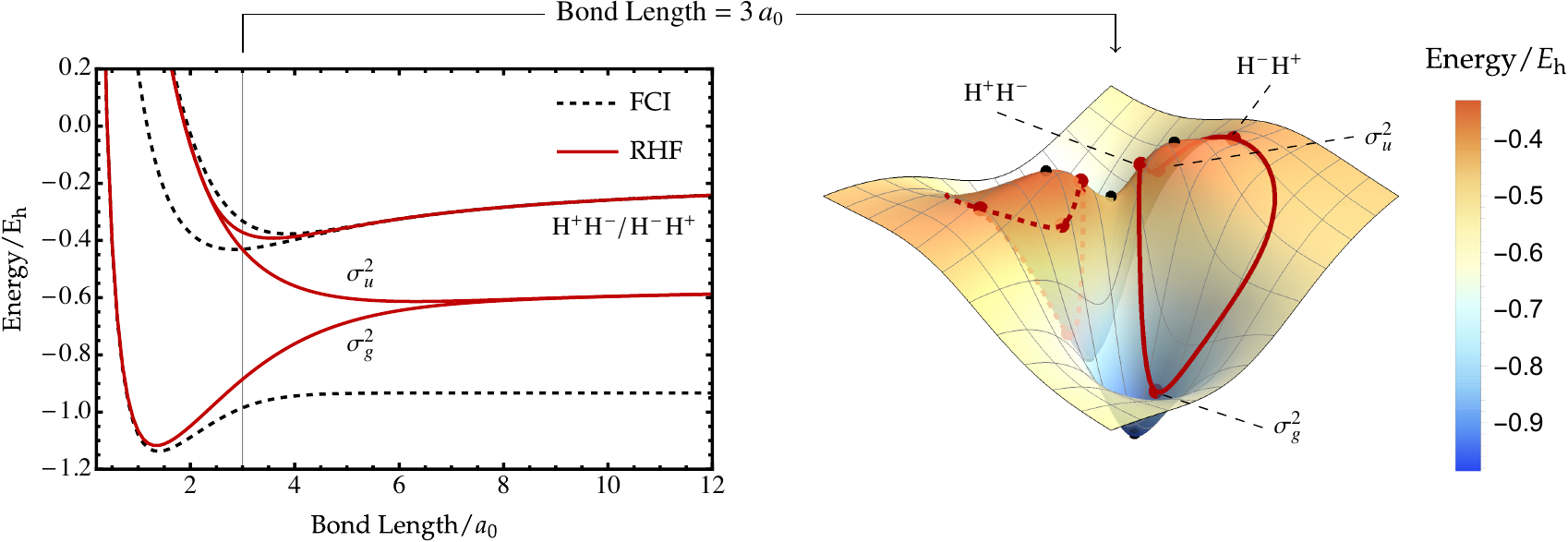}
\caption{The space of possible RHF wave functions forms a one-dimensional submanifold 
(right panel; red curve) on the exact singlet energy surface for \ce{H2} (STO-3G) with a bond length of $\SI{3}{\bohr}$.
Multiple RHF solutions (left panel) correspond to constrained stationary points on the RHF manifold
(right panel; red dots).
\edit{The sign-permuted RHF wave functions are denoted by a dashed red curve (red panel).}
}
\label{fig:fci_rhf_Topology}
\end{figure*}

Any normalised wave function approximation $\ket*{\tilde{\Psi}(\bm{t})}$ with variational parameters $\bt$ 
can be represented as a point on the exact hypersphere using a linear expansion in the many-particle basis
\begin{equation}
\ket*{\tilde{\Psi}(\bm{t})} = \sum_{I=1}^{\Nh} \tilde{c}_I(\bm{t}) \ket{\Phi_I}.
\end{equation}
Like the exact wave function, this expansion is invariant to the particular choice of orthogonal basis determinants.
Since the number of parameters $t_i$ is generally smaller than the Hilbert space size, 
the approximate wave functions form a constrained submanifold of the exact hypersphere.
The structure of this approximate  submanifold is implicitly defined  by the mathematical form of the parametrisation.
Geometrically, the approximate energy is  given as
\begin{equation}
E(\bt) = \tilde{\bc}(\bm{t})^\T\, \bH\, \tilde{\bc}(\bm{t}),
\end{equation}
and the constrained local gradient is defined as
\begin{equation}
\qty(\widetilde{\grad} E)_i 
= 
\frac{\partial E(\bt)}{\partial t_i} = 2\,  \qty( \frac{\partial\tilde{\bc}(\bm{t})}{\partial t_i})^\T \bH \, \tilde{\bc}(\bm{t}).
\label{eq:localGradientTerms}
\end{equation}
Here, the partial derivatives of the coefficient vector define the local 
tangent basis of the approximate submanifold, representing the ket vectors
\begin{equation}
\ket*{\eta_i} = \pdv{t_i}\ket*{\tilde{\Psi}(\bm{t})}.
\label{eq:TangetVectors}
\end{equation}
In analogy with the exact wave function, the approximate local  gradient
corresponds to the global gradient [Eq.~\eqref{eq:GlobalGradient}] projected into the space 
spanned by the approximate tangent vectors [Eq.~\eqref{eq:TangetVectors}].
Optimal stationary points of the approximate energy then occur 
when $\widetilde{\grad} E = \bm{0}$.

The HF approach illustrates how well-known approximate stationary 
conditions can be recovered with this geometric perspective.
Although HF theory is usually presented as an iterative self-consistent approach,\cite{Roothaan1951,Hall1951}  
the HF wave function can also be parametrised using an exponential transformation of a reference determinant to give\cite{Thouless1960,Douady1980}
\begin{equation}
\ket{\Phi(\bm{\kappa})} = \exp(\hat{\kappa})\ket{\Phi_0}.
\end{equation}
Here, $\hat{\kappa}$ is a unitary operator constructed from closed-shell single excitations and de-excitations, 
represented in second-quantisation as\cite{HelgakerBook}
\begin{equation}
\hat{\kappa} = \sum_{ai} \kappa_{ai} \qty(
a_{a}^{\dagger} a_{i}^{\vphantom{\dagger}} - a_{i}^{\dagger} a_{a}^{\vphantom{\dagger}}
)
\label{eq:singlesExp}
\end{equation}
where $\kappa_{ai}$ are the variable parameters.
The tangent vectors at $\bm{\kappa} = \bm{0}$ correspond to the singly-excited configurations, i.e.
\begin{equation}
\ket{\eta_{ai}} 
= \left. \frac{\partial\ket{\Phi(\bm{\kappa})}}{\kappa_{ai}} \right\rvert_{\bm{\kappa}=\bm{0}}   
= \ket{\Phi_i^a}.
\end{equation}
These approximate tangent vectors span a subspace of the exact tangent space on the full hypersphere.
Using Eq.~\eqref{eq:localGradientTerms}, the local HF gradient is then given by\cite{Douady1980,Voorhis2002} 
\begin{equation}
\qty(\widetilde{\grad} E)_{ai}  
= 2 \mel*{\Phi_i^a}{\cH}{\Phi_0} = 2 F_{ai}, 
\end{equation}
which corresponds to twice the virtual-occupied components of the Fock matrix in the MO basis.
As expected, the stationary condition $\widetilde{\grad} E =\bm{0}$ recovers Brillouin's theorem for HF
convergence.\cite{SzaboBook}

\subsection{Multiple Hartree--Fock solutions}
\label{subsec:MultipleSolutions}

\begin{figure}[b!]
\includegraphics[width=\linewidth]{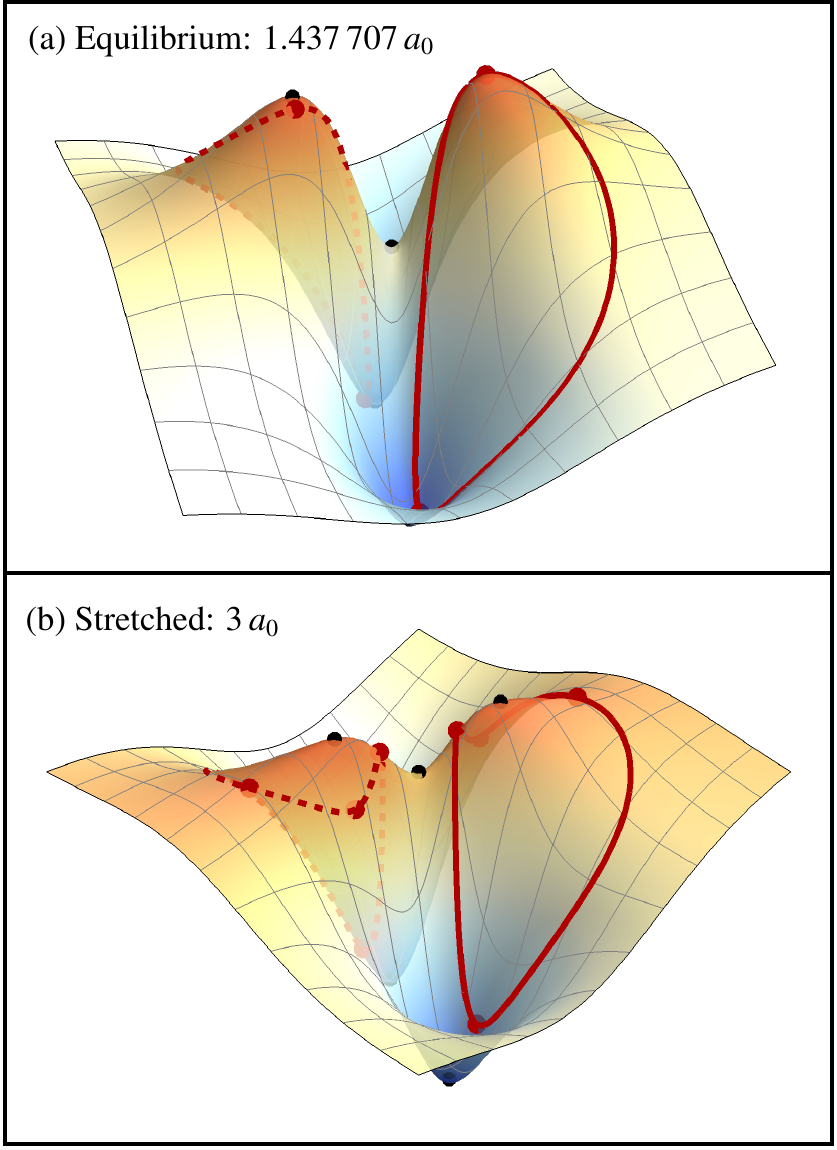}
\caption{Mapping between the RHF submanifold (solid red line) and the exact 
singlet energy surface for \ce{H2} (STO-3G) at (a) the equilibrium bond length $\SI{1.437707}{\bohr}$ and 
(b) a stretched geometry $\SI{3}{\bohr}$.
\edit{Sign-permuted RHF wave functions are denoted by dashed red curves.}
}
\label{fig:fci_rhf_topo_comparison}
\end{figure}

Although the HF wave function has fewer parameters than the exact wave function, 
there can be more HF stationary points than exact eigenstates.\cite{Stanton1968,Burton2018}
For example, in dissociated \ce{H2} with a minimal basis set, there are four closed-shell restricted HF (RHF) solutions
and only three exact singlet states.\cite{Burton2018}
In contrast to the exact eigenstates, there can also be multiple HF solutions with the same Hessian index, 
although this index generally increases with energy.\cite{Burton2018}
Furthermore, HF solutions do not necessarily exist for all molecular geometries and can disappear at so-called 
``Coulson--Fischer'' points.\cite{Coulson1949,Slater1951,Thom2008,Huynh2019,Burton2018,Burton2021}
These phenomena can all be understood through the geometric mapping between the approximate HF 
submanifold and the exact energy landscape.

Consider the RHF approximation for the singlet states of \ce{H2} (STO-3G).
Only two RHF solutions exist at the equilibrium geometry, while an additional higher-energy 
pair of degenerate solutions emerge in the dissociation limit, as shown in the left panel of Fig.~\ref{fig:fci_rhf_Topology} 
(see Ref.~\onlinecite{Burton2018} for further details).
In this system, the RHF submanifold forms a continuous one-dimensional subspace of the exact 
energy surface, illustrated by the red curve in Fig.~\ref{fig:fci_rhf_Topology} (right panel). 
This submanifold includes \emph{all possible} closed-shell Slater determinants for the system
and is fixed by the wave function parametrisation.
Approximate solutions then correspond to constrained stationary points of the energy along
 the RHF submanifold, which occur when the global energy gradient has no component 
parallel to the red curve.
The existence  and properties of these solutions is completely determined by the mapping between
the RHF submanifold and the exact energy landscape.

At bond lengths near the equilibrium structure of \ce{H2} (STO\nobreakdash-3G), the RHF submanifold extends relatively close to the exact 
global minimum and maximum, as illustrated in Fig.~\ref{fig:fci_rhf_topo_comparison}a.
This mapping results in only two constrained stationary 
points, the global minimum and maximum of the RHF energy, which correspond to
the symmetry-pure $\sigg^2$ and $\sigu^2$ configurations respectively. 
Notably, the RHF $\sigu^2$ global maximum represents a doubly-excited
state that cannot be accurately described by TD-HF or CIS.\cite{Burke2005}
However, the RHF submanifold cannot get sufficiently close to the exact open-shell 
singlet state to provide a good approximation, and there is no stationary point representing this single excitation.

\begin{figure}[b!]
\includegraphics[width=\linewidth]{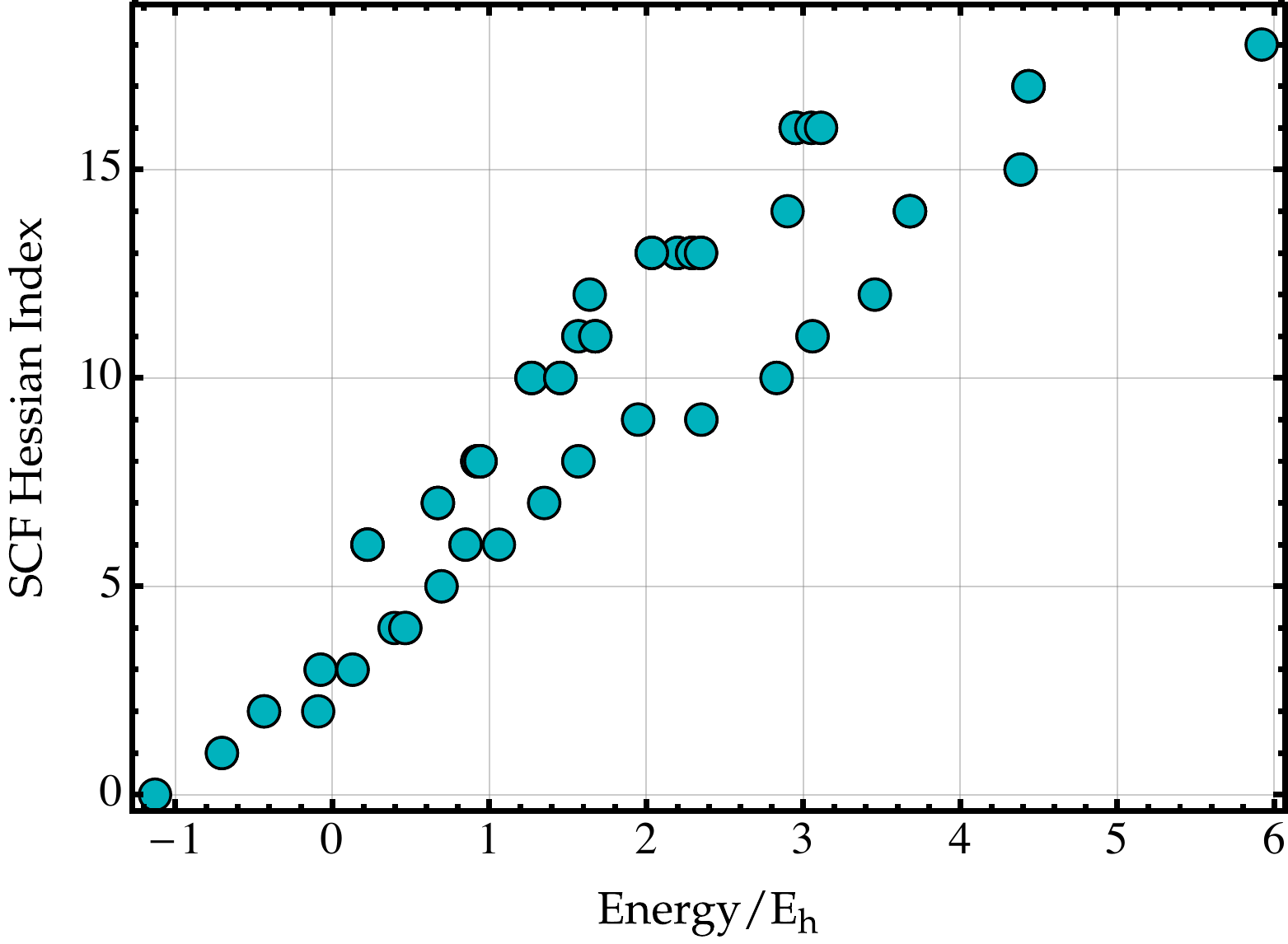}
\caption{Comparison of the energy and SCF Hessian index computed for the  orbital-optimised excited configurations
of \ce{H2} (cc-pVDZ\cite{Dunning1989}) at $R=\SI{1.437707}{\bohr}$ using  unrestricted HF.}
\label{fig:HessCompare} 
\end{figure}

As the \ce{H-H} bond is stretched towards dissociation, the exact energy landscape on the 
hypersphere changes while the RHF submanifold remains fixed by the functional form of the approximate wave function. 
Therefore, the exact energy evolves \textit{underneath} the RHF submanifold, creating changes in the approximate energy 
that alter the properties of the constrained stationary points.
For example, at a sufficiently large bond length, the RHF submanifold no longer 
provides an accurate approximation to the exact global maximum and instead encircles it  to give two local 
maxima and a higher-energy local minimum of the RHF energy (Fig.~\ref{fig:fci_rhf_topo_comparison}b).
These local maxima represent the spatially-symmetry-broken  RHF solutions that tend towards to the ionic 
dissociation limit (left panel in Fig.~\ref{fig:fci_rhf_Topology}), while the higher-energy local minimum represents the $\sigu^2$  configuration.
Combined with the global minimum, this gives a total of four RHF solutions, 
in contrast to only three exact singlet states.
Consequently, we find that the number of RHF solutions can exceed the number of exact eigenstates 
because the RHF submanifold is a highly constrained non-linear subspace of the exact energy landscape. 
Furthermore, it is the structure of this constrained subspace that creates a high-energy local minimum
at dissociation, while the exact energy landscape has no local minima at any geometry.

\begin{figure*}[htb!]
\includegraphics[width=0.8\linewidth]{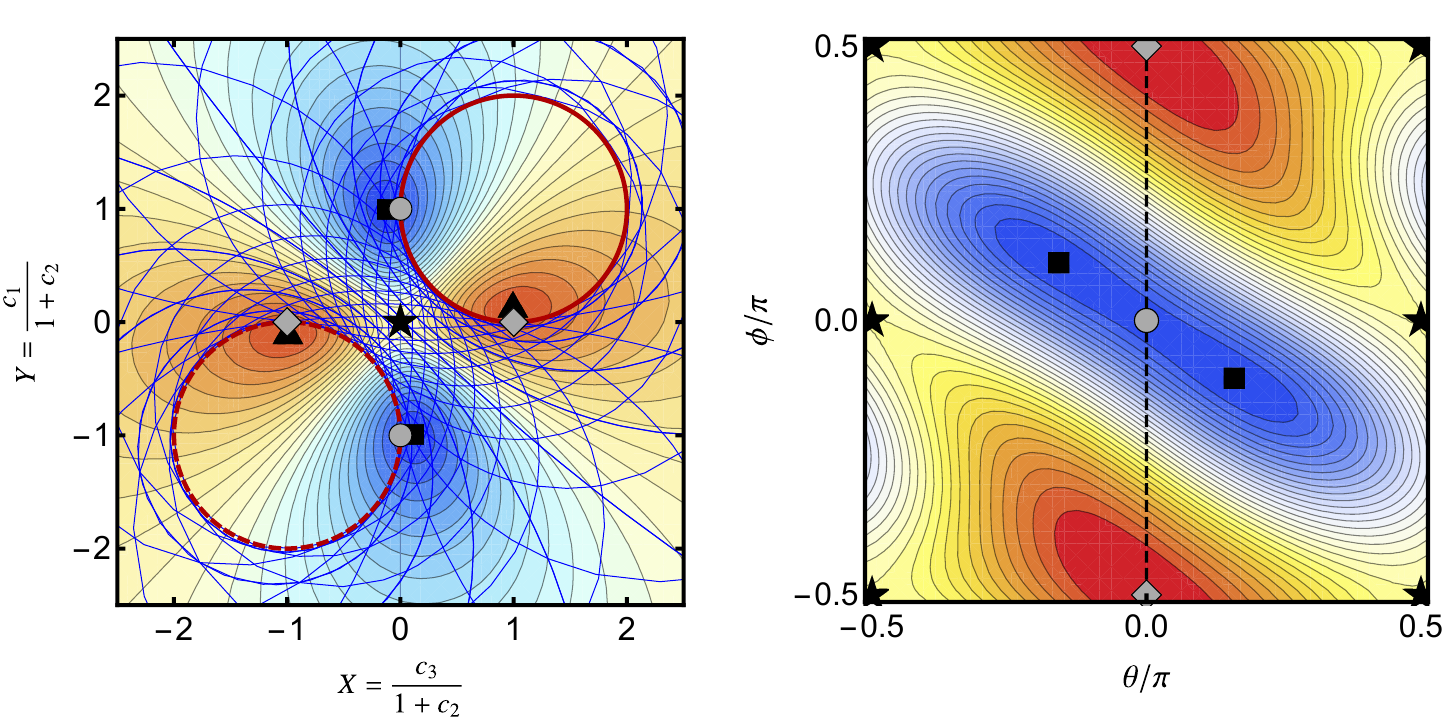}
\caption{
\textit{Left:} RHF (solid and dashed red curves) and ESMF (blue mesh) constrained submanifolds superimposed on a
stereographic projection [see Eq.~\eqref{eq:stereo}] 
of the exact singlet energy landscape for \ce{H2} (STO-3G) at a bond length 
of $R=\SI{1.437707}{\bohr}$.
Exact stationary points and RHF solutions occur at the black and red dots respectively. 
\textit{Right:}
The ESMF energy for singlet \ce{H2} (STO-3G) as a function of the wave function parameters 
[see Eq.~\eqref{eq:OOcisH2}].
Black and grey symbols indicate stationary points of the ESMF or RHF energy respectively, and the RHF submanifold corresponds to
$\theta = 0$ (dashed black line).}
\label{fig:ooCIS}
\end{figure*}

Finally, 
for real-valued HF wave functions, the Hessian of the approximate energy is computed using the second derivatives\cite{Seeger1977}
\begin{equation}
\frac{\partial^2 E(\bt) }{\partial t_{ia} \partial t_{jb}} = 2\, \Big(\mel*{\Phi_{i}^{a}}{\cH - E}{\Phi_{j}^{b}} +  \mel*{\Phi}{\cH - E}{\Phi_{ij}^{ab}} \Big),
\label{eq:OrbHess}
\end{equation}
where now open-shell orbital rotations are now allowed.
To investigate how the HF Hessian index changes with energy, 
the unrestricted excited configurations obtained from the ground-state RHF orbitals  of 
\ce{H2} (cc-pVDZ\cite{Dunning1989}) at
$R~=~\SI{1.437707}{\angstrom}$ were optimised using the initial maximum overlap method\cite{Barca2018} 
in Q-Chem 5.4.\cite{QChem54}
The corresponding Hessian indices are plotted against the optimised energy in Fig.~\ref{fig:HessCompare}.
Similarly to the exact eigenstates, there is a general increase in the Hessian index at higher energies.
However, unlike the exact eigenstates, the approximate Hessian index does not increase monotonically with the energy.
These results strengthen the conclusion of Ref.~\onlinecite{Burton2021} that approximate HF excited states are generally
higher-index saddle points of the energy.

\subsection{Excited-state mean-field theory}
\label{subsec:CISandPCCD}

While multiple HF solutions are relatively well understood, the energy landscape of orbital-optimised post-HF 
wave functions remains less explored.
The simplest excited-state extension of HF theory is the CIS wave function,\cite{Foresman1992,Dreuw2005} 
constructed as a linear combination of singly-excited determinants.
\edit{CIS generally provides a qualitatively correct description of singly excited states,
but it is less reliable for multiconfigurational or charge transfer excitations.\cite{Dreuw2005} 
The systematic overestimate of CIS for charge transfer excitations can be attributed to the absence of 
orbital relaxation effects.\cite{Subotnik2011}}
Therefore, to improve this description, the orbitals and CI coefficients can be 
simultaneously optimised to give the state-specific excited-state mean-field (ESMF) wave function, defined for singlet 
states as\cite{Shea2018}
\begin{equation}
\ket{\Psi_{\text{ESMF}}} = \exp(\hat{\kappa}) \qty[c_0 
+ \frac{1}{\sqrt{2}} \sum_{ia} c_{ia} 
\qty( a^{\dagger}_{a} a^{\vphantom{\dagger}}_{i} 
+  a^{\dagger}_{\bar{a}} a^{\vphantom{\dagger}}_{\bar{i}}  ) ] \ket{\Phi_0} .
\end{equation}
Here, the reference determinant is retained in the expansion and orbital rotations
are parametrised by the unitary rotation defined in
Eq.~\eqref{eq:singlesExp}.
In recent years, efficient optimisation of this \textit{ansatz} has been 
successfully applied to charge transfer and core excitations.\cite{Shea2018,Zhao2020a,Hardikar2020,Garner2020} 
Alternative approaches to include orbital relaxation effects in CIS using perturbative corrections
have also been investigated.\cite{Liu2012,Liu2013,Liu2014}

Geometrically, the single excitations for any closed-shell reference determinant
define the tangent vectors to the RHF submanifold. 
Therefore, the orbital-optimised ESMF submanifold contains the RHF wave functions and 
all points that lie in the combined tangent spaces of the RHF submanifold, 
as illustrated for the singlet states of \ce{H2} (STO-3G) in Fig.~\ref{fig:ooCIS} (left panel).
Like multiple RHF solutions, state-specific ESMF solutions correspond to stationary 
points of the energy constrained to the ESMF wave function manifold. 

The ESMF wave function for singlet \ce{H2} (STO-3G) can be constructed by parametrising the occupied 
and virtual molecular orbitals with a single rotation angle $\phi$ as  
\begin{subequations}
\begin{align}
\psi_1(\br)  &= \cos \phi\, \sigg(\br) + \sin \phi\, \sigu(\br),
\\
\psi_2(\br)  &= \cos \phi\, \sigu(\br) -\sin \phi\, \sigg(\br).
\end{align}
\end{subequations}
and defining the normalised singlet CI expansion with a second rotation angle $\theta$ to give
\begin{equation}
\ket{\Psi_{\text{ESMF}}} = \cos \theta\, \abs*{\psi_1  \bar{\psi}_1  } + \sin \theta \, \frac{\abs*{\psi_1  \bar{\psi}_2  } + \abs*{\psi_2  \bar{\psi}_1 } }{\sqrt{2}} .
\label{eq:OOcisH2}
\end{equation} 
The corresponding energy landscape at the equilibrium bond length $R=\SI{1.437707}{\bohr}$ is 
shown in Fig.~\ref{fig:ooCIS} (right panel) as a function of $\theta$ and $\phi$, 
with the RHF approximation indicated by the dashed black line at $\theta=0$.
Stationary points representing the exact open-shell singlet state occur at $(\theta, \phi) = (\pm\frac{\pi}{2},0)$ and 
$(\pm \frac{\pi}{2},\pm \frac{\pi}{2})$, denoted by \edit{black stars}, with optimal orbitals that correspond to $\sigg(\br)$ and $\sigu(\br)$.
In common with the exact energy landscape, these open-shell singlet solutions form index-1 
saddle points of the singlet energy.
On the other hand, the local maximum representing the double excitation has no contribution from the single excitations and 
reduces to the closed-shell $\sigu^2$ RHF solution \edit{(grey diamonds)}.
This lack of improvement beyond RHF can be understood from the structure of the singlet ESMF submanifold:
the exact double excitation is encircled by the RHF submanifold and therefore cannot be reached by any of the 
tangent spaces to the RHF wave function (Fig.~\ref{fig:ooCIS}: left panel).

However, the most surprising observation from Fig.~\ref{fig:ooCIS} is not  the higher-energy 
stationary points of the  ESMF energy, but the location of the global minimum.
Counter-intuitively, the RHF ground state becomes an index-1 saddle point of the ESMF energy \edit{(grey circles)} and there is a lower-energy 
solution at  $(\theta, \phi) = (\pm 0.5026,\mp 0.3304)$ that corresponds to the exact ground state \edit{(black squares)}.
This lower-energy solution occurs because rotating the orbitals away from an optimal HF solution breaks 
Brillioun's condition and introduces new coupling terms between the reference and singly-excited configurations
that allow the energy to be lowered below the RHF minimum.
Since the RHF ground state is stationary with respect to both orbital rotations and the introduction of single excitations, 
this cooperative effect can only occur when the orbital and CI coefficients are optimised simultaneously.
Furthermore, the combined orbital and CI Hessian is required to diagnose the RHF ground state as a saddle point of the ESMF energy, 
highlighting the importance of considering the full parametrised energy landscape.

The existence of an ESMF global minimum below the RHF ground state challenges 
the idea that CIS-based wave functions are only useful for approximating excited states.
From a practical perspective, current ESMF calculations generally underestimate excitation energies because the
multiconfigurational wave function used for the excited state can capture some electron correlation, 
while the single-determinant RHF ground state remains completely uncorrelated.\cite{Shea2018}
\edit{Although CIS is often described as an uncorrelated excited-state theory, the wave function
is inherently multi-configurational and becomes correlated 
when the first-order density matrix is not idempotent.\cite{Surjan2007}
These circumstances generally correspond to excitations with more than one dominant 
natural transition orbital.\cite{Plasser2016}
Therefore, it is not too surprising that the ESMF global minimum can provide a 
correlated representation of the ground state.}
As a result, using state-specific ESMF wave functions for both the ground and excited states
should provide a more balanced description of an electronic excitation.

\begin{figure}[b!]
\setlength\fboxrule{1pt}
\includegraphics[width=\linewidth]{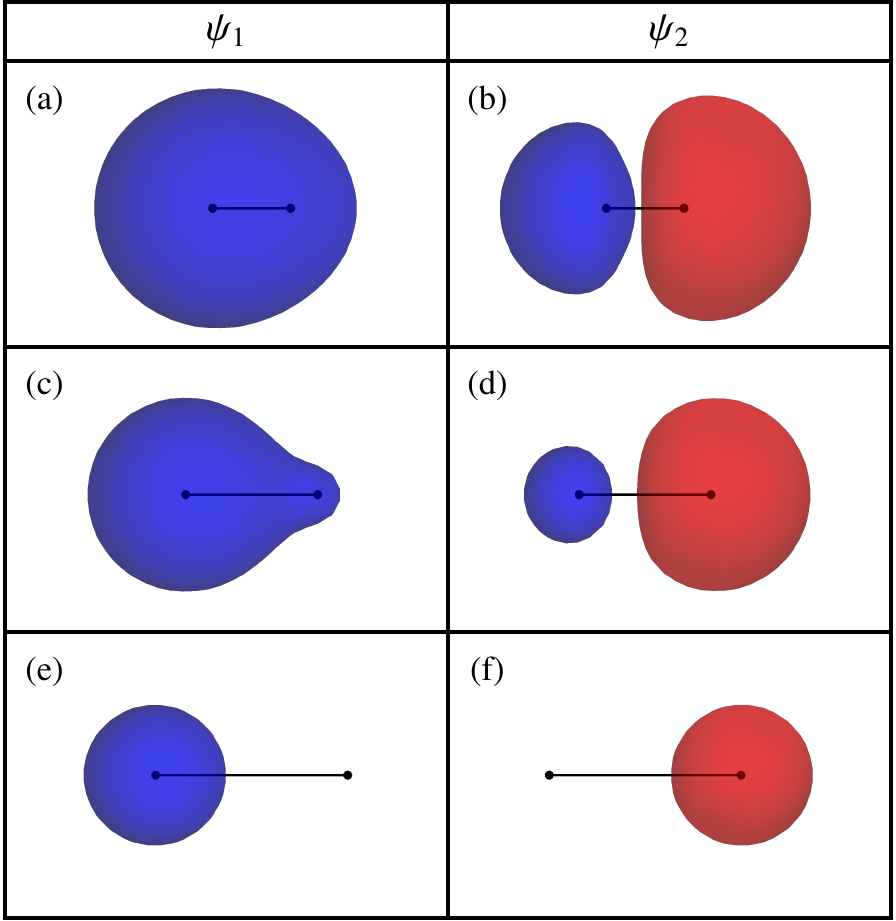}
\caption{%
\edit{%
Spatial orbitals for the optimised ESMF ground state of \ce{H2} (STO-3G) at bond lengths of 
(a)--(b) $\SI{1.437707}{\bohr}$,
(c)--(d) $\SI{3.0}{\bohr}$, and
(e)--(f) $\SI{6.0}{\bohr}$,
plotted with an isosurface value of $\pm0.05$.
The reference configuration corresponds to $|\psi_1 \bar{\psi}_1|$ with the ESMF wave function defined in Eq.~\eqref{eq:OOcisH2}.}
}
\label{fig:esmf_orbs}
\end{figure}

Since the global minimum is exact across all \ce{H2} bond lengths,
orbital-optimised ESMF may also provide an alternative reference wave function for capturing 
static correlation in single-bond breaking processes.
\edit{The success of this \textit{ansatz} can be compared to the spin-flip CIS (SF-CIS) approach, where the 
CI expansion is constructed using spin-flipping excitations from a high-spin reference determinant.\cite{Krylov2001}
SF-CIS gives an accurate description of the \ce{H2} ground state because single spin-flip excitations
from an open-shell $\sigg \sigu$ reference produce both the $\sigg^2$ and $\sigu^2$ configurations.
In contrast, the ESMF global minimum contains a closed-shell reference determinant 
with symmetry-broken orbitals, as shown in Fig.~\ref{fig:esmf_orbs}.
These orbitals resemble the $\sigg$ and $\sigu$ MOs at short geometries [Fig.~\ref{fig:esmf_orbs}(a)--(b)] where the reference determinant $|\psi_1 \bar{\psi}_1|$
dominates the ESMF wave function.
As the bond is stretched, the optimised orbitals localise on opposite H atoms to give an ionic reference
state [Fig.~\ref{fig:esmf_orbs}(e)--(f)]  and the single  excitations correspond to the localised configurations required for the diradical ground state.
}

\begin{figure*}
\setlength\fboxrule{1pt}
\includegraphics[width=\linewidth]{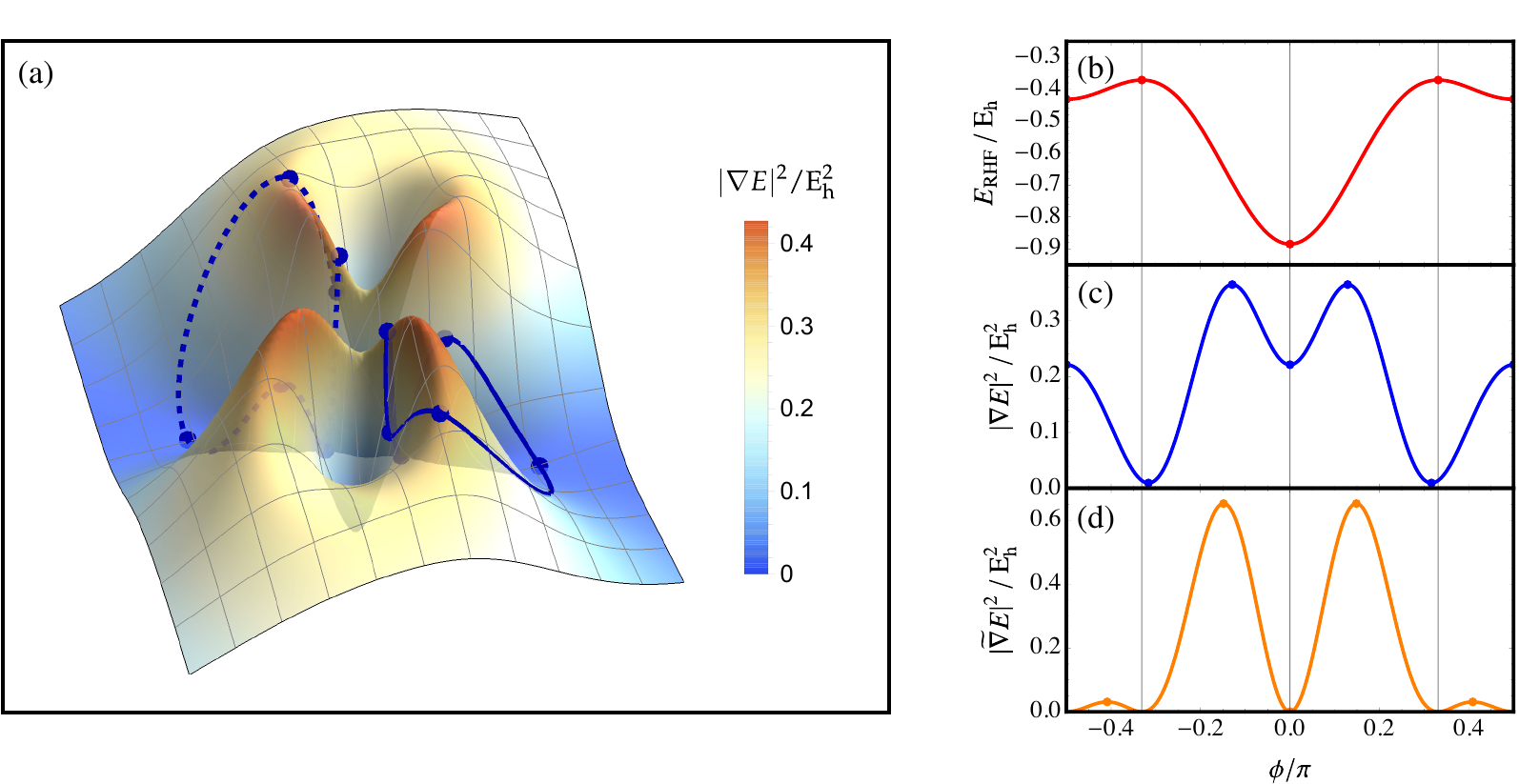}
\caption{(a) Exact square-gradient landscape for singlet \ce{H2} (STO-3G) at a bond length of $\SI{3}{\bohr}$. 
Restricted $\sigma$-SCF solutions identified using variance optimisation are equivalent to
constrained stationary points (blue dots) on the single-determinant subspace (blue curve) and its sign-related
copy (dashed blue curve).
(b) RHF energy as a function of the single orbital rotation angle $\phi$ [see Eq.~\eqref{eq:orbParam}]. 
(c) Exact square-gradient of RHF wave functions with stationary points corresponding to $\sigma$-SCF solutions.
(d) Square-magnitude of the constrained RHF energy gradient, with minima corresponding to RHF energy stationary points.
}
\label{fig:fci_rhf_variance}
\end{figure*}

\subsection{Unphysical solutions in variance optimisation}
\label{subsec:VarianceChallenges}

Approximate state-specific variance minimisation can also be considered as an optimisation of 
the exact variance constrained to the approximate wave function submanifold. 
Variance minimisation is increasingly being applied to target  excited states because it 
turns the optimisation of higher-energy stationary points into a minimisation
problem\cite{Messmer1969,Shea2017,PinedaFlores2019,Ye2017,Ye2019,David2021} 
and is easily applied for correlated wave functions using VMC.\cite{Otis2020,Cuzzocrea2020}
In practice, these algorithms often use the folded-spectrum objective function\cite{Wang1994}
\begin{equation}
\Omega = \frac{\mel*{\tilde{\Psi}}{(\cH - \omega)^2}{\tilde{\Psi}}}{\braket*{\tilde{\Psi}}{\tilde{\Psi}}}
\label{eq:omegaTarget}
\end{equation}
to target an excited state with an energy near $\omega$, before self-consistently updating
$\omega$ until a variance stationary point is reached with $\omega =E$.\cite{Ye2017,Ye2019,Cuzzocrea2020}
However, the structure of the variance landscape for approximate wave functions, and the properties 
of its stationary points, are relatively unexplored.

Since the variance is equivalent to the exact square-gradient $\abs{\grad E}^2$, this landscape
can be investigated using the mapping between an approximate wave function and the exact 
square-gradient landscape derived in Sec.~\ref{subsec:SquaredGradientVariance}.
Consider the RHF approximation in \ce{H2} (STO-3G) with the 
doubly-occupied orbital parametrised as
\begin{equation}
\psi(\bm{r}) = \cos \phi \, \sigg(\bm{r}) + \sin \phi \,\sigu(\bm{r}).
\label{eq:orbParam}
\end{equation}
Variance optimisation for HF wave functions has been developed by Ye \etal\ as the iterative $\sigma$-SCF 
method, which uses a variance-based analogue of the Fock matrix.\cite{Ye2017,Ye2019}
In Figure~\ref{fig:fci_rhf_variance}a, the RHF submanifold is shown as a subspace of the 
exact singlet square-gradient landscape at $R=\SI{3}{\bohr}$, with
optimal $\sigma$-SCF solutions corresponding to the constrained stationary points.
Since the RHF approximation cannot reach any exact eigenstates in this system, the square-gradient is non-zero 
for all RHF wave functions and there is no guarantee that the $\sigma$-SCF solutions
coincide with stationary points of the constrained energy.
This feature is illustrated in Figs.~\ref{fig:fci_rhf_variance}b and \ref{fig:fci_rhf_variance}c, where the energy and 
square-gradient are compared for the occupied orbital defined in Eq.~\eqref{eq:orbParam}.
There are three constrained minima of the square-gradient for these RHF wave functions.
The first, at $\phi=0$ corresponds to the $\sigg^2$ RHF ground state, while the other two represent 
ionic configurations that are similar, but not identical, to the local maxima of the RHF energy.\cite{Ye2017}
However, the $\sigu^2$ configuration, which forms a local minimum of the RHF energy, becomes a constrained
local maximum of the energy square-gradient.

\begin{figure}[t!]
\includegraphics[width=\linewidth]{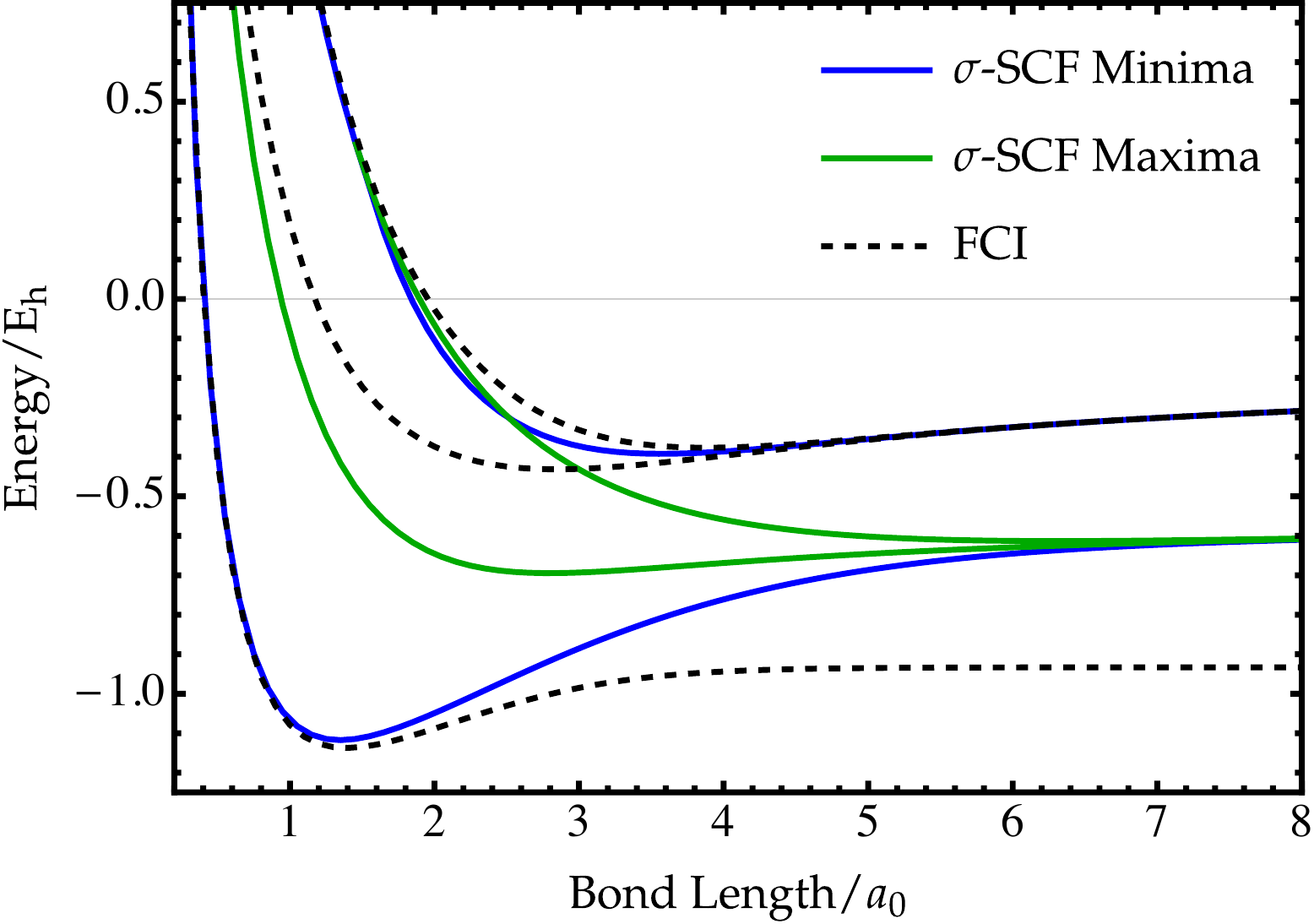}
\caption{Energy of restricted $\sigma$-SCF solutions corresponding 
to constrained stationary points of the variance in \ce{H2} (STO-3G), 
including local minima (blue) and maxima (green).
Exact singlet energies are shown for comparison (black dashed).}
\label{fig:h2_sigmaSCF}
\end{figure}

Energies corresponding to the complete set of RHF variance stationary points 
for \ce{H2} (STO-3G) are shown in Fig.~\ref{fig:h2_sigmaSCF}, with variance minima and maxima denoted 
by blue and green curves respectively.
These solutions closely mirror the low-energy $\sigma$-SCF states identified using the 3-21G basis set in Ref.~\onlinecite{Ye2017}.
Despite becoming a local maximum of the variance at large bond lengths, the $\sigma$-SCF approach
identifies the $\sigu^2$ solution at all geometries.\cite{Ye2017}
Therefore, iterative methods such as $\sigma$-SCF must be capable of converging onto higher-index stationary points 
of the variance.
However, not all higher-index stationary points correspond to physically meaningful solutions. 
For example, an additional degenerate pair of $\sigma$-SCF maxima can be found at all bond lengths in \ce{H2}, 
with an energy that lies between the $\sigg^2$ and $\sigu^2$ solutions. 
These unphysical maxima exist where the 
RHF manifold passes over the square-gradient barriers created by non-stationary points on the
exact energy landscape (see Fig.~\ref{fig:fci_rhf_variance}a).
The high non-convexity of the exact square-gradient landscape means that these unphysical 
higher-index stationary points of the variance are likely to be very common for approximate wave functions.

As an alternative to variance minimisation, excited state-specific wave functions can 
be identified by \edit{directly searching for points where the approximate local gradient 
becomes zero $\widetilde{\grad}E = \bm{0}$.
For example, Shea and Neuscamman introduced an approach that modifies the objective function Eq.~\eqref{eq:omegaTarget}
using Lagrange multipliers to
ensure that the approximate local gradient vanishes at a solution.\cite{Shea2018}
Alternatively, Hait and Head-Gordon proposed the square-gradient minimisation (SGM) algorithm that directly minimises
the square-magnitude of the local electronic gradient $\abs*{\widetilde{\grad}E}^2$.\cite{Hait2020}
While the SGM approach is closely related to  minimising the exact variance,} all stationary points 
of the approximate energy now become minima of local square-gradient 
with $\abs*{\widetilde{\grad}E}^2=0$, as shown for the RHF wave functions of \ce{H2} (STO-3G) in Fig.~\ref{fig:fci_rhf_variance}d. 
Unphysical square-gradient stationary points, such as those resembling the RHF variance 
maxima in Fig.~\ref{fig:fci_rhf_variance}c, can then be easily identified with $\abs*{\widetilde{\grad}E}^2 \neq 0$.
However, since the approximate energy can have more stationary points than the exact energy, there will 
be more non-stationary points of $\abs*{\widetilde{\grad}E}^2$ corresponding to inflection points between 
stationary states of the approximate energy (\textit{cf}.\ Fig.~\ref{fig:fci_rhf_variance}a and \ref{fig:fci_rhf_variance}d).
These additional non-stationary points will make the approximate $\abs*{\widetilde{\grad}E}^2$ landscape even
more non-convex than the constrained variance landscape, which can make numerical 
optimisation increasingly difficult.

\section{Implications for Optimisation Algorithms }
\label{sec:Discussion}
 

\edit{We have seen that the exact energy landscape for real wave functions has only two minima, 
corresponding to negative and positive sign-permutations of the exact ground state.
In addition, there is only one sign-permuted pair of  index-$k$ saddle points corresponding to the $k$-th excited state.
On the contrary, approximate methods may have higher-energy local minima and multiple index-$k$ saddle points, 
although the maximum Hessian index is dictated by the number of approximate wave function parameters.
These higher-energy local minima result from the wave function constraints introduced by lower-dimensional approximations.
}

The structure of the exact energy landscape highlights the importance of developing excited state-specific
algorithms that can converge onto arbitrary saddle points of the energy. 
Any type of stationary point can be identified using 
iterative techniques that modify the SCF procedure,
including the maximum overlap method\cite{Gilbert2008,Besley2009} 
and state-targeted energy projection.\cite{CarterFenk2020} 
In addition, modified quasi-Newton optimisation 
of the SCF energy\cite{Levi2020,Levi2020a} should perform well,
while state-specific CASSCF\cite{Tran2019,Tran2020} 
may also benefit from similar second-order optimisation.
Alternatively, modified eigenvector-following may allow excited states with a particular Hessian 
index to be targeted.\cite{Doye2002,Wales2003,Burton2021}
However, methods that search for local minima of the 
energy, including SCF metadynamics\cite{Thom2008} and 
direct minimisation,\cite{Voorhis2002} will perform less well 
for excited states and are better suited to locating the global minimum or symmetry-broken solutions.

\edit{In addition, the structure of the energy square-gradient landscape elucidates 
the challenges faced by variance optimisation approaches.
Each exact eigenstate forms a minimum of the variance and minima adjacent in energy are separated
by an index-1 saddle point with height proportional to the square of the corresponding energy difference.}

\edit{Low barriers on the exact variance landscape offer a new perspective on the 
convergence drift observed in excited-state VMC calculations.\cite{Cuzzocrea2020,Otis2020} 
In Ref.~\onlinecite{Cuzzocrea2020}, variance optimisation was found to 
drift away from the intended target state defined by the initial guess, 
passing through eigenstates sequentially in energy until converging onto the state with the lowest variance. 
By definition, a deterministic minimisation algorithm cannot climb over a barrier to escape a variance minimum.
However, the statistical uncertainty of stochastic VMC calculations means that they \textit{can} climb a variance 
barrier if the height is sufficiently low.
Essentially, the variance barrier becomes ``hidden'' by statistical noise.
The index-1 variance saddle points connecting states adjacent in energy may then 
explain why the optimisation drifts through eigenstates sequentially in energy\cite{Cuzzocrea2020} and
why this issue is more prevalent in systems with small energy gaps.\cite{Otis2020}
Alternatively, the VMC wave function may  simply not extend far 
enough into the exact variance basin of attraction of the target state to create an approximate local minimum.
However, the use of highly-sophisticated wave functions in Ref.~\onlinecite{Cuzzocrea2020} would suggest that 
this latter explanation is unlikely.
}

\edit{An additional concern is the presence of unphysical local variance minima or higher-index stationary 
points that occur when the exact variance is constrained to an approximate wave function manifold.
For minimisation algorithms such as VMC or generalised variational principles, the presence of many 
higher-index saddle points may increase the difficulty of convergence. 
There is also a risk of getting stuck in a spurious local minimum on the constrained manifold which does 
not correspond to a physical minimum on the exact variance landscape.
In contrast, iterative self-consistent algorithms such as $\sigma$-SCF\cite{Ye2017,Ye2019}  are capable of converging onto 
higher-index stationary points and it may be difficult to establish the physicality of these solutions.
Therefore, iterative variance optimisation requires careful analysis to ensure the physicality of solutions, 
for example by using sufficiently accurate initial guesses.
}

\edit{Finally, although not considered in this work, the generalised variational principle developed by Neuscamman and co-workers  includes Lagrange multipliers 
to simultaneously optimise multiple objective functions that target a particular excited eigenstate.\cite{Shea2018,Shea2020,Hanscam2021}
The combined Lagrangian may include functionals that target a particular energy (such as Eq.~\eqref{eq:omegaTarget}), 
the square-magnitude of the local gradient, orthogonality to a nearby state, or a desirable 
dipole moment.\cite{Hanscam2021}
This approach is particularly suited to systems where there is a good initial guess for the excited state or its properties,
and a good choice of objective functions can significantly accelerate numerical convergence.\cite{Shea2020}
These constraints may also help to prevent the drift of stochastic variance optimisation algorithms by
increasing the barrier heights between minima with the correct target properties.
}


\section{Concluding Remarks}
\label{sec:ConcludingRemarks}
This contribution has introduced \edit{a geometric perspective on the energy landscape}
of exact and approximate state-specific electronic structure theory.
In this framework, exact ground and excited states become stationary points of an energy landscape 
constrained to the surface of a unit hypersphere while approximate wave functions form constrained subspaces.
Furthermore, the Hamiltonian variance  $\mel*{\Psi}{(\cH -E)^2}{\Psi}$ is directly proportional 
to the square-magnitude of the exact energy gradient. 
Deriving this geometric framework allows exact and approximate excited state-specific methods to 
be investigated on an equal footing and leads to the following key results:
\begin{enumerate}[itemsep=0em]
\item{Exact excited states form saddle points of the energy with the number of 
downhill directions equal to the excitation level;}
\item{The local energy and second derivatives at an exact stationary point 
can be used to deduce the entire energy spectrum of a system;}
\item{\edit{The exact energy landscape has only two minima, corresponding to sign-permutations
 of the ground state;}}
\item{Approximate excited solutions are generally saddle points of the energy 
and their Hessian index increases with the excitation energy;}
\item{Physical minima of the variance are separated from states adjacent in energy by index-1 saddle points. 
The barrier height is proportional to the square of the energy difference between the two states;}
\end{enumerate}
\edit{While only the simple \ce{H2} example has been considered, these results are sufficient to establish a set of guiding principles
for developing robust optimisation algorithms for state-specific excitations.
Future work will investigate how fermionic anti-symmetry affects the relationship between exact and approximate electronic energy landscapes 
for systems with multiple same-spin electrons.}


Beyond state-specific excitations, the exact energy landscape may also provide a new perspective for understanding
the broader properties of wave function approximations.
For example, this work has shown that the orbital-optimised ESMF
wave function can describe the exact ground state of dissociated \ce{H2} (STO-3G) for all bond lengths
using only the reference determinant and single excitations.
This observation suggests that the orbital-optimised ESMF ground state may provide an 
alternative black-box wave function for capturing static correlation in single-bond dissociation.
Alternatively, drawing analogies between a Taylor series approximation on the exact energy landscape 
and second-order perturbation theory may provide an orbital-free perspective on the divergence 
of perturbative methods for strongly correlated systems.
Finally, investigating how more advanced methods such as multi-configurational SCF,\cite{RoosBook} variational CC,\cite{Cooper2010,Marie2021a} or Jastrow-modified antisymmetric geminal power\cite{Neuscamman2011,Neuscamman2013} 
approximate exact stationary points on the energy landscape may inspire 
entirely new ground- and excited-state wave function approaches.

\section*{Acknowledgements}
H.G.A.B. was financially supported by New College, Oxford through the Astor Junior Research Fellowship.
The author owes many thanks to Antoine Marie, Alex Thom, David Wales, David Tew, and Pierre-Fran\c{c}ois Loos for discussions
and support throughout the development of this work.
The author is also thanks the Reviewers for insightful comments that have improved this work.

\section*{References}
\bibliography{manuscript}

\end{document}